\documentclass[prd,twocolumn,showpacs,amsmath,amssymb,floatfix,nofootinbib]{revtex4}
\usepackage{graphicx,color,dcolumn,booktabs}

\begin{document}
\title{Study on the radiative decays of $\Upsilon(nS)\to \eta_b+\gamma$}

\author{Hong-Wei Ke$^1$}\email{khw020056@hotmail.com}
\author{Xue-Qian Li$^2$}\email{lixq@nankai.edu.cn}
\author{Xiang Liu$^{3,4}$}\email{xiangliu@lzu.edu.cn}\affiliation{$^1$School of Science, Tianjin University, Tianjin 300072, China\\
$^2$School of Physics, Nankai University, Tianjin 300071, China\\
$^3$School of Physical Science and Technology, Lanzhou University, Lanzhou 730000, China\\
$^4$Research Center for Hadron and CSR Physics,
Lanzhou University $\&$ Institute of Modern Physics of CAS, Lanzhou 730000, China}

\date{\today}
\begin{abstract}
\noindent In this work, we investigate the characteristics of the
spin-singlet state $\eta_b$ of the bottomonia family via the
radiative decays of $\Upsilon(nS)\to \eta_b+\gamma$. The theoretical
estimation of the decay widths is carried out in terms of the
light-front quark model (LFQM). Recently CLEO and BaBar
collaborations have measured
$\mathcal{B}(\Upsilon(3S)\rightarrow\gamma\eta_b)$ and the mass of
${\eta_b}$. In terms of the data we fix the concerned input
parameters in our calculations of $\Upsilon(nS)\to \eta_b+\gamma$. A
special attention is paid on the transition of $\Upsilon(5S)\to
\eta_b+\gamma$. The BELLE data showed that the width of
$\Upsilon(5S)\to \Upsilon(2S,1S)+\pi\pi$ is two orders larger than
that of $\Upsilon(4S)\to \Upsilon(2S,1S)+\pi\pi$, thus some
theoretical explanations have been proposed. Among them, it is
suggested the inelastic final state interaction (IFSI)
$\Upsilon(5S)\to B\bar B\to \Upsilon(1S)+\pi\pi$ may be a natural
one. If so, a similar mechanism also applies to $\Upsilon(5S)\to
B^{(*)}\bar B^{(*)}\to \eta_b+\gamma$, the precise measurement
would serve as a good test whether $\Upsilon(5S)$ possess exotic
components. Our calculation in the LFQM indicates that the rate of
the direct process $\Upsilon(5S)\rightarrow\eta_b+\gamma$ is not
anomalous compared to $\Upsilon(mS)\rightarrow\eta_b+\gamma\,
(m=1,2,3,4)$, thus if the IFSI does apply, the rate of
$\Upsilon(5S)\rightarrow\eta_b+\gamma$ should be larger than the
others by orders.
\end{abstract}

\pacs{13.25.Gv, 13.30.Ce, 12.39.Ki}
\maketitle

\section{introduction}

Since the relativistic and higher-order $\alpha_s$ corrections are
less important for bottomonia than for any other $q\bar q$ systems,
study on bottomonia may offer more direct information about the
hadron configuration and perturbative QCD. The spin-triplet state of
bottomonia $\Upsilon(nS)$ and the P-states $\chi_b(nP)$ were
discovered decades ago, however its partner the singlet state
$\eta_b$ evaded detection for long time, even though much efforts
were made. Many phenomenological researches on $\eta_b$ have been
done by some
groups \cite{Hao:2007rb,Ebert:2002pp,Motyka:1997di,Liao,Recksiegel:2003fm,Gray:2005ur,Eichten:1994gt,Ke:2007ih}.
Different approaches result in different level splitting $\Delta
M=\Upsilon(1S)-\eta_b(1S)$.  In Ref. \cite{Recksiegel:2003fm} the
authors used an improved perturbative QCD approach to get $\Delta
M=44$ MeV; using the potential model suggested
in \cite{Buchmuller:1980su} Eichten and Quigg estimated $\Delta M=87$
MeV \cite{Eichten:1994gt}; in Ref. \cite{Motyka:1997di} the authors
selected a non-relativistic Hamiltonian with spin dependent
corrections to study the spectra of heavy quarkonia and got $\Delta
M$=57 MeV; the lattice prediction is $\Delta M$=51 MeV \cite{Liao},
whereas the lattice result calculated in Ref. \cite{Gray:2005ur} was
$\Delta M=64\pm14$MeV. Ebert $et\, al.$ \cite{Ebert:2002pp} directly
studied spectra of heavy quarkonia in the relativistic quark model
and gave $m_{\eta_b}=9.400$ GeV.

The Babar Collaboration \cite{:2008vj} first measured
$\mathcal{B}(\Upsilon(3S)\rightarrow\gamma\eta_b)=(4.8\pm0.5\pm0.6)\times10^{-4}$,
$M(\eta_b)=9388.9^{+3.1}_{-2.3}\pm2.7$ MeV and $\Delta M=
71.4^{+3.1}_{-2.3}\pm2.7$ MeV in 2008 and new data were released in
2009\cite{:2009pz}\footnote{We thank Dr. Michael Roney for telling
us the discovery history of $\eta_b$ as the Babar collaboration
published its 10-sigma discovery of the $\eta_b$ in 2008  whereas in
2009 the CLEO collaboration released a re-analysis of their data and
reported a 4-sigma confirmation of the Babar discovery.}. More
recently the CLEO Collaboration \cite{Bonvicini:2009hs} confirmed
the observation of $\eta_b$ using the database of 6 million
$\Upsilon(3S)$ decays and assuming $\Gamma(\eta_b)\approx$10 MeV,
they obtained
$\mathcal{B}(\Upsilon(3S)\rightarrow\gamma\eta_b)=(7.1\pm1.8\pm1.1)\times10^{-4}$,
$M_{\eta_b}=9391.8\pm6.6\pm2.0$ MeV and the hyperfine splitting
$\Delta M= 68.5\pm6.6\pm2.0$ MeV, whereas using the database with 9
million $\Upsilon(2S)$ decays they obtained
$\mathcal{B}(\Upsilon(2S)\rightarrow\gamma\eta_b)<8.4\times10^{-4}$
at 90\% confidential level It is noted that the data of the two
Collaborations are in accordance on $M_{\eta_b}$, but the central
values of $\mathcal{B}(\Upsilon(3S)\rightarrow\gamma\eta_b)$ are
different. However, if the experimental errors are taken into
account, the difference is still within one standard deviation.

Some theoretical
work \cite{Radford:2009qi,Colangelo:2009pb,Seth:2009ba} is devoted
to account the experimental results.

In Ref. \cite{Ebert:2002pp} the authors studied these radiative
decays and estimated
$\mathcal{B}(\Upsilon(3S)\rightarrow\eta_b+\gamma)=4\times10^{-4}$,
$\mathcal{B}(\Upsilon(2S)\rightarrow\eta_b+\gamma)=1.5\times10^{-4}$
and
$\mathcal{B}(\Upsilon(1S)\rightarrow\eta_b+\gamma)=1.1\times10^{-4}$
with the mass $m_{\eta_b}$ = $9.400$ GeV. Their results about
$m_{\eta_b}$ and $\mathcal{B}(\Upsilon(3S)\rightarrow\eta_b+\gamma)$
are close to the data. The authors of Ref. \cite{Choi:2007se}
systematically investigated the magnetic dipole transition
$V\rightarrow P\gamma$ in the light-front quark
model (LFQM) \cite{Jaus:1999zv,Cheng:2003sm,Hwang:2006cua,Wei:2009nc}. In
the QCD-motivated effective Hamiltonian  there are several free
parameters, i.e., the quark mass and $\beta$ in the wavefunction (the
notation of $\beta$ was given in the aforementioned literatures)
which are fixed by the variational principle, then
$\mathcal{B}(\Upsilon(1S)\rightarrow\eta_b+\gamma)$ was calculated
and the central value is $8.4\,({\rm or}\,7.7)\times
10^{-4}$. \footnote{The different values correspond to the deferent
potentials adopted in calculations.}

\begin{center}
\begin{figure}[htb]
\begin{tabular}{c}
\scalebox{0.9}{\includegraphics{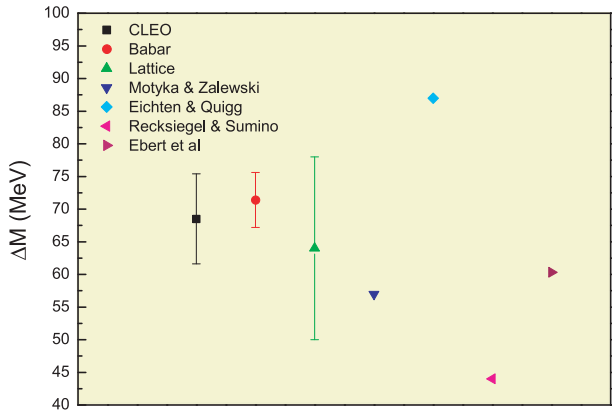}}
\end{tabular}
\caption{$\Delta M$ coming from different experimental measurement and
theoretical work.}\label{DM}
\end{figure}
\end{center}

It is also noted that the mass of $m_{\eta_b}=9.657\,({\rm or}
\,9.295)$ GeV presented in Ref.\cite{Choi:2007se} deviates from the
data ($m_{\eta_b}=9391.8\pm6.6\pm2.0$ MeV) and the fitted $\beta$
values are different for singlet and triplet \cite{Cheng:2003sm}.
Here we are going to take an alternative way to fix the values of
$\beta$.

Since experimentally, $m_{\eta_b}$ is determined by
$\mathcal{B}(\Upsilon(nS)\rightarrow\eta_b+\gamma)$ and a study on
the radiative decays  can offer us much information about the
characteristics of $\eta_b$, one should carefully investigate the
transition within a more reliable theoretical framework. That is
the aim of the present work, namely we will use the LFQM to evaluate the hadronic matrix element which
is governed by the non-perturbative QCD. The method is proven to
be successful for calculating the transition rates of the
processes where light hadrons exist in the final states.

In this work, we first fix $\beta_n$'s for $\Upsilon(nS)$ in terms
of their decay constants. Then, using the data of
$\mathcal{B}(\Upsilon(3S)\rightarrow\eta_b+\gamma)$ and $m_{\eta_b}$,
we determine $\beta_{\eta_b}$. With the parameters being fixed, we
are able to estimate the rates
$\mathcal{B}(\Upsilon(1S)\rightarrow\eta_b+\gamma)$,
$\mathcal{B}(\Upsilon(2S)\rightarrow\eta_b+\gamma)$,
$\mathcal{B}(\Upsilon(4S)\rightarrow\eta_b+\gamma)$ and
$\mathcal{B}(\Upsilon(5S)\rightarrow\eta_b+\gamma)$. Since
$\mathcal{B}(\Upsilon(1S)\rightarrow\eta_b+\gamma)$ is sensitive to
$\Delta M$, the measurement of
$\mathcal{B}(\Upsilon(1S)\rightarrow\eta_b+\gamma)$ would be helpful
for accurately determining the mass of $\eta_b$.

Recently the transition rates of
$\Upsilon(5S)\rightarrow\Upsilon(1S,2S)+\pi\pi$ were measured by the
BELLE Collaboration \cite{Belle-1} and it was found that the widths
exceed by more than two orders of magnitude the previously measured
partial widths between lower $\Upsilon$ resonances. The authors
\cite{Chao-1} suggested that the re-scattering processes of
$\Upsilon(5S)\to  B^{(*)}\bar B^{(*)}\to
\Upsilon(mS)+\sigma/f_0(980)\to \Upsilon(mS)+\pi\pi$ make
substantial contributions to the observable rate of the dipion
transition of $\Upsilon(5S)$ because its mass exceeds the production
threshold of $B^{(*)}\bar B^{(*)}$, so that the intermediate bosons
$B^{(*)}$ and $\bar B^{(*)}$ are on their mass-shell. They apply the
same mechanism to study the transition of $\Upsilon(5S)\to
\Upsilon(1S)+\eta$ \cite{Chao-2} and find that the re-scattering
processes would enhance its width by almost two orders of magnitude.
There are different interpretations for the anomalous enhancement of
the $\Upsilon(5S)$ decays that the measured resonance
$\Upsilon(10870)$ is a mixture of $b\bar b$ bound state in the $5S$
state with a hybrid $b\bar bg$ or a tetraquark $b\bar bq\bar
q$ \cite{Hou}.

Thus, we would like to further test the mechanism in the radiative
decays of $\Upsilon(5S)$. In the radiative decay of
$\Upsilon(5S)\to\eta_b+\gamma$, the re-scattering processes
$\Upsilon(5S)\to B^{(*)}\bar B^{(*)}\to \Upsilon(mS)+\gamma$ also
exist and one only needs to replace the effective vertex of
$B^{(*)}\bar B^{(*)}\Upsilon(1S)$ and $B^{(*)}\bar B^{(*)}\eta$ by
the electromagnetic vertex $B^{(*)}\bar B^{(*)}\gamma$ and
$B^{(*)}\bar B^{(*)}\eta_b$ respectively in the diagrams given in
Ref. \cite{Chao-1}. Thus one can expect that the corresponding
mechanism should enhance the ratio of $\Upsilon(5S)\to
\eta_b+\gamma$. Our calculations show that the theoretical
estimation on the enhancement factor strong depends on the parameter
$g_{B^{(*)}B^{(*)}\eta_b}$ (see below for more details). The future
measurements on the radiative decays of $\Upsilon(10870)$ can help
to determine if it is the $\Upsilon(5S)$ state as long as the rate
of $\Upsilon(5S)\to \eta_b+\gamma$ is obviously larger than that of
lower resonances of the family, otherwise other mechanisms may be
more favored. Anyhow, the radiative decays would provide a decisive
probe for the re-scattering mechanism.

This paper is organized as follows: after the introduction, in
section II we present our calculations of the form factors for
$V\rightarrow P\gamma$ in the LFQM
and the corresponding numerical results. In the section III we
study the possible re-scattering effects on $\Upsilon(5S)\to \eta_b+
\gamma$. The section IV is devoted to our conclusion and
discussion.

\section{$\Upsilon(nS)\to \eta_b +\gamma$ in the LFQM}
\subsection{Description of $\Upsilon(nS)\to \eta_b +\gamma$ in the LFQM}

\begin{center}
\begin{figure}[htb]
\begin{tabular}{c}
\scalebox{0.9}{\includegraphics{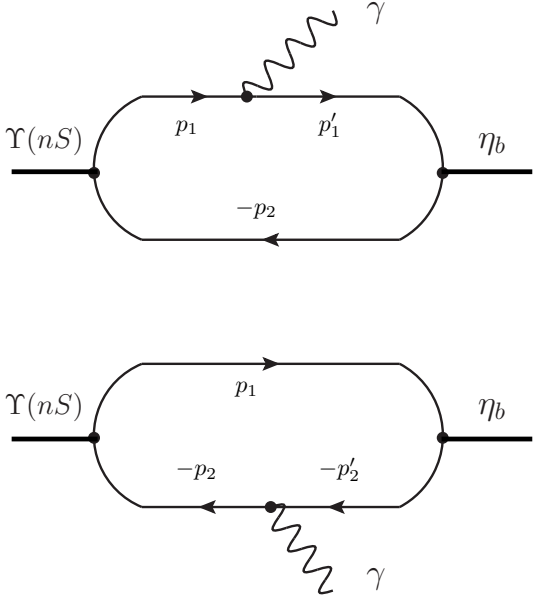}}
\end{tabular}
\caption{Feynman diagrams depicting the radiative decay $\Upsilon(nS)\to \eta_b+\gamma$.\label{fig:LFQM}}
\end{figure}
\end{center}

The Feynman diagrams describing $\Upsilon(nS)\to \eta_b+ \gamma$
are presented in Fig. \ref{fig:LFQM}. In this work, we calculate
the transition rate of the radiative decays $\Upsilon(nS)\to
\eta_b +\gamma$ in the LFQM.

The transition amplitude of $\Upsilon(nS)\to \eta_b+ \gamma$ can be
expressed in terms of the form factor $\mathcal{
F}_{\Upsilon(nS)\to\eta_b}(q^2)$ which is defined as
\cite{Choi:2007se,Hwang:2006cua}
\begin{eqnarray}\label{2S1}
&&\langle
\eta_b(\mathcal{P}')|J_{em}^\mu|\Upsilon(\mathcal{P},h)\rangle\nonumber\\&&=ie\,\varepsilon^{\mu\nu\rho\sigma}\epsilon_\nu(\mathcal{P},h)q_\rho
\mathcal{P}_\sigma\mathcal{ F}_{\Upsilon(nS) \to\eta_b}(q^2),
\end{eqnarray}
where $\mathcal{P}$  and $\mathcal{P}'$ are the four-momenta of
$\Upsilon(nS)$ and $\eta_b$. $q=\mathcal{P}-\mathcal{P}'$ is the
four-momentum of the emitted photon and
$\epsilon_\nu(\mathcal{P},h)$ denotes the polarization vector of
$\Upsilon(nS)$ with helicity $h$. For applying the LFQM, we first
let the photon be virtual, i.e. leave its mass-shell $q^2=0$ into
the un-physical region of $q^2<0$. Then $\mathcal{ F}_{\Upsilon(nS)\to
\eta_b}(q^2)$ can be obtained in the $q^+=0$ frame with $q^2=q^+q^-
- {\bf q}^2_\perp=-{\bf q}^2_\perp<0$. Then we just analytically
extrapolate $\mathcal{ F}_{\Upsilon(nS) \to\eta_b}({\bf q}^2_\perp)$
from the space-like region to the time-like region ($q^2\geq 0$). By
taking the limit $q^2\rightarrow 0$, one obtains $\mathcal{
F}_{\Upsilon(nS)\to \eta_b}( q^2=0)$.

By means of the light front quark model, one can obtain the
expression of form factor $\mathcal{ F}_{\Upsilon(nS)\to \eta_b}(q^2)$
\cite{Choi:2007se}:
\begin{eqnarray}\label{21}
\mathcal{ F}_{\Upsilon(nS)\to\eta_b}(q^2)= e_bI(m_1,m_2,q^2) + e_{b}
I(m_2,m_1,q^2),
\end{eqnarray}
where $e_{b}$ is the electrical charge for bottom quark ,
$m_1=m_2=m_b$ and
\begin{eqnarray}\label{22}
I(m_1,m_2,q^2) &=&\int^1_0 \frac{dx}{8\pi^3}\int d^2{\bf k}_\perp
\frac{\phi(x, {\bf k'}_\perp)\phi(x,{\bf k}_\perp)}
{x_1\tilde{M_0}\tilde{M'_0}}
\nonumber\\
&&\times \biggl\{{\cal A} + \frac{2} {{\cal M}_0} [{\bf k}^2_\perp -
\frac{({\bf k}_\perp\cdot{\bf q}_\perp)^2}{{\bf q}^2_\perp}]
\biggr\}.
\nonumber\\
\end{eqnarray}
${\cal A}=x_2m_1+x_1m_2$, $x=x_1$ and the other variables in Eq.
(\ref{22}) are defined in Appendix. In the covariant light-front
quark model the authors of Ref. \cite{Hwang:2006cua} obtained the
same  form factor $\mathcal{ F}_{\Upsilon(nS)\to
\eta_b}(\mathbf{q}^2)$. The decay width for
$\Upsilon(nS)\rightarrow \eta_b+\gamma$ is easily achieved
\begin{eqnarray}\label{23}
\Gamma(\Upsilon(nS)\rightarrow
\eta_b+\gamma)=\frac{\alpha}{3}\bigg[\frac{m_{\Upsilon(nS)}^2-m_{\eta_b}^2}{2m_{\Upsilon(nS)}}\bigg]^3
\mathcal{ F}^2_{\Upsilon(nS) \to\eta_b}(0),
\end{eqnarray}
where $\alpha$ is the fine-structure constant and
$m_{\Upsilon(nS)},\; m_{\eta_b}$ are the masses of $\Upsilon(nS)$
 and $\eta_b$ respectively.

\subsection{Numerical results}

In Ref.\cite{Choi:2007se} the authors
fixed the parameter $\beta_{b\bar b}=1.145$ or 1.803 \footnote{The
authors of Ref. \cite{Choi:2007se} used the harmonic oscillator and
linear potential forms for the confinement term in their
computations, thus they obtained two different values for the
$\beta$ parameter. Since later in our calculations we do not
evaluate the spectra of the concerned hadrons, the concrete value of
$\beta$ does not influence our numerical results.} in the Gaussian
wavefunction using the variational method with $m_b=5.2$ GeV.
However their predictions on the mass $m_{\eta_b}=9.657\,({\rm or}
\,9.295)$ GeV and $m_{\Upsilon(1S)}=9.691\,({\rm or} \,9.558$) GeV (the values in the brackets are obtained with a different potential
form), which obviously deviate from data ($m_{\eta_b}=9300\pm 20\pm
20$ GeV and $m_{\Upsilon(1s)}=9460.30\pm 0.26$ GeV \cite{PDG08}), thus
we are going to fix $\beta$ in an alternative way.

First we set the b quark mass as $m_b=4.64$ GeV which was used in
Ref. \cite{Cheng:2003sm}. Then we extract  the decay constant
$f_{\Upsilon(nS)}$ of $\Upsilon(nS)$  from the data
$\Gamma(\Upsilon(nS)\rightarrow e^+e^- )$ \cite{PDG08} with
\begin{eqnarray}\label{31}
\Gamma(V\rightarrow
e^+e^-)=\frac{4\pi}{27}\frac{\alpha^2}{M_V}f^2_{V}.
\end{eqnarray}
In parallel,  we can calculate the constant in the LFQM using the
formula \cite{Jaus:1999zv,Cheng:2003sm}
\begin{eqnarray}\label{31}
f_V&=&\frac{\sqrt{N_c}}{4\pi^3M}\int dx\int
d^2k_\perp\frac{\varphi}{\sqrt{2x(1-x)}\tilde M_0}
\biggl[xM_0^2\nonumber\\&&-m_1(m_1-m_2)-k^2_\perp+\frac{m_1+m_2}{M_0+m_1+m_2}k^2_\perp\biggl],\nonumber\\
\end{eqnarray}
equating the two results. $\beta_{\Upsilon(nS)}$ is determined (see
Table \ref{tab:etab1}). Then, we use the formula presented in
Section II to calculate
$\mathcal{B}(\Upsilon(3S)\rightarrow\eta_b+\gamma)$ and compare it
with the central value of the experimental data
$\mathcal{B}(\Upsilon(3S)\rightarrow\eta_b+\gamma)=(7.1\pm1.8\pm1.1)\times10^{-4}$ \cite{Bonvicini:2009hs}
or
$\mathcal{B}(\Upsilon(3S)\rightarrow\eta_b+\gamma)=(4.8\pm0.5\pm0.6)\times10^{-4}$ \cite{:2009pz}
to fit $\beta_{\eta_b}$ and the corresponding values are presented
in Table \ref{tab:etab2}. At last with all the parameters we
estimate $\mathcal{B}(\Upsilon(1S)\rightarrow\gamma\eta_b)$,
$\mathcal{B}(\Upsilon(2S)\rightarrow\eta_b+\gamma)$,
$\mathcal{B}(\Upsilon(4S)\rightarrow\eta_b+\gamma)$ and
$\mathcal{B}(\Upsilon(5S)\rightarrow\eta_b+\gamma)$ which are shown
in Table \ref{tab:etab2}. It is noted that at this step, we only
consider the direct decay modes, but for
$\Upsilon(5S)\rightarrow\eta_b+\gamma$, the re-scattering effect may
play a dominant role as mentioned in the introduction and we will
discuss the details in next section.

In order to illustrate the dependence of our results on $m_b$, we
re-set $m_b=5.2$ GeV, which was adopted by the authors of
Ref. \cite{Choi:2007se} and fitted $\beta_{\Upsilon(nS)}$ and
$\beta_{\eta_b}$ again. Using the new parameter the
$\mathcal{B}(\Upsilon(nS)\rightarrow\eta_b+\gamma)$ is computed and
the result is also listed in Table \ref{tab:etab2}. From the
Table \ref{tab:etab2}, we find

1. The predicted branching ratios
$\mathcal{B}(\Upsilon(1S)\rightarrow\eta_b+\gamma)$ and
$\mathcal{B}(\Upsilon(4S)\rightarrow\eta_b+\gamma)$  are not
sensitive to  $m_b$ and $\beta$, but
$\mathcal{B}(\Upsilon(2S)\rightarrow\eta_b+\gamma)$ and
$\mathcal{B}(\Upsilon(5S)\rightarrow\eta_b+\gamma)$ slightly change
as $m_b$ and $\beta$ vary within certain ranges;

2. Our results about
$\mathcal{B}(\Upsilon(1S)\rightarrow\eta_b+\gamma)$  and
$\mathcal{B}(\Upsilon(2S)\rightarrow\eta_b+\gamma)$ are somehow
larger than that given in Ref. \cite{Ebert:2002pp};

3. If the final state interaction is not taken into account, the
branching ratio of $\Upsilon(5S)\rightarrow\eta_b+\gamma$ is not
anomalous compared to $\Upsilon(mS)\rightarrow\eta_b+\gamma\,
(m=1,2,3,4)$. Because the BELLE Collaboration \cite{Abe:2007tk} found
the rate of
$\Upsilon(5S)\rightarrow\Upsilon(1S,2S)+\pi\pi$ is
anomalously large compared to the similar dipion transitions between
lower $\Upsilon$ resonances, one has reason to doubt if such anomaly
would appear in $\Upsilon(5S)\rightarrow\eta_b+\gamma$.

\begin{figure}
\begin{center}
\begin{tabular}{cc}
\includegraphics[width=9cm]{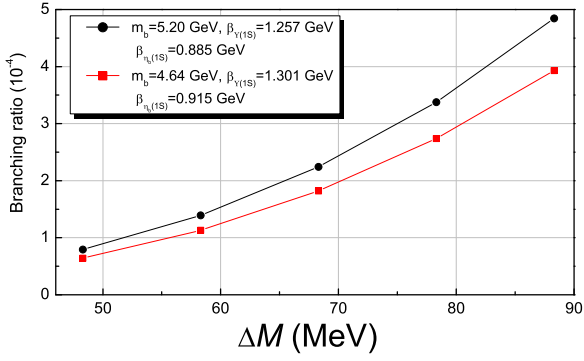}
\end{tabular}
\end{center}
\caption{The dependance of
$\mathcal{B}(\Upsilon(1S)\rightarrow\gamma\eta_b)$ on $\Delta M$.}
\label{fig:LFQM2}
\end{figure}

It is noted that $\mathcal{B}(\Upsilon(1S)\rightarrow\eta_b+\gamma)$
is sensitive to $m_{\eta_b}$ (or\, $\Delta M$) since the decay width is
proportional to $(\Delta M)^3$, thus as $\Delta M$ is small, i.e.
the masses of initial and daughter mesons are close to each other,
any small changes of $m_{\eta_b}$ which has not been accurately
measured yet \cite{PDG08}, can lead to a remarkable difference. In
Fig. \ref{fig:LFQM2} we display the dependence of
$\mathcal{B}(\Upsilon(1S)\rightarrow\eta_b+\gamma)$ on $\Delta M$.
Thus the accurate measurement on
$\mathcal{B}(\Upsilon(1S)\rightarrow\eta_b+\gamma)$ will be a great
help to determine the mass of $m_{\eta_b}$.

\begin{widetext}\begin{center}
\begin{table}
\caption{ The $b$ quark mass and the parameters $\beta$ in the wavefunction in the unit of GeV.} \label{tab:etab1}
\begin{tabular}{c|ccccc}\toprule[1pt]
$m_b$  & $\beta_{\Upsilon(1S)}$  & $\beta_{\Upsilon(2S)}$&
$\beta_{\Upsilon(3S)}$&$\beta_{\Upsilon(4S)}$&$\beta_{\Upsilon(5S)}$
\\\midrule[1pt]
4.64&1.301&0.924&0.795&0.657&0.673
\\ 5.20&1.257&0.894&0.769&0.635&0.652\\\bottomrule[1pt]
\end{tabular}
\end{table}

\begin{table}
\caption{ The branching ratio of
$\Upsilon(nS)\rightarrow\gamma\eta_b$.}
\label{tab:etab2}
\begin{tabular}{c|c|cccc}\toprule[1pt]
$m_b$    & $\beta_{\eta_b}$ &
$\mathcal{B}{(\Upsilon(1S)\rightarrow\eta_b+\gamma)}$&$\mathcal{B}{(\Upsilon(2S)\rightarrow\eta_b+
\gamma)}$&$\mathcal{B}{(\Upsilon(4S)\rightarrow\eta_b+\gamma)}$
&$\mathcal{B}{(\Upsilon(5S)\rightarrow\eta_b+\gamma)}$\\\midrule[1pt]
4.64&0.915\footnote{fitted from CLEO
data}&2.25$\times10^{-4}$&4.91$\times 10^{-4}$&4.66$\times
10^{-6}$&9.27$\times10^{-8}$\\
& 0.901\footnote{fitted from BaBar
data}&2.52$\times10^{-4}$&9.21$\times 10^{-4}$&3.86$\times
10^{-6}$&6.97$\times10^{-8}$
\\\hline
5.20&0.885$^a$&1.83$\times10^{-4}$&2.27$\times10^{-4}$&4.31$\times10^{-6}$&8.57$\times10^{-8}$
\\  & 0.874$^b$& 2.05$\times 10^{-4}$&5.31$\times10^{-4}$&3.53$\times 10^{-6}$&6.32$\times10^{-8}$\\
\bottomrule[1pt]
\end{tabular}
\end{table}\end{center}
\end{widetext}

\section{Possible re-scattering effects inducing a large $\mathcal{B}(\Upsilon(5S)\to \eta_b +\gamma)$ }

\begin{center}
\begin{figure}[htb]
\begin{tabular}{c}
\scalebox{0.45}{\includegraphics{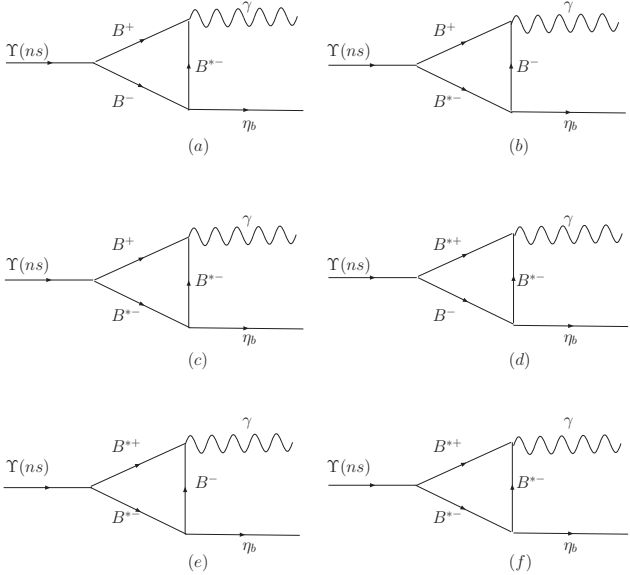}}
\end{tabular}
\caption{The  diagrams for $\Upsilon(nS)\rightarrow
B^{(*)+}B^{(*)-}\to \eta_b+\gamma$. Other diagrams can be obtained by
a charge conjugation transformation $B^{(*)}\rightleftarrows \bar{B}^{ (*)}$. \label{fig:fsi}}
\end{figure}
\end{center}

As aforementioned, the re-scattering of hadrons may remarkably
enhance the rate of $\Upsilon(nS)\,(n>4)\rightarrow \eta_b+ \gamma$.
Similar to the re-scattering effects on in the branching ratios of
$\Upsilon(5S)\rightarrow \Upsilon(1S)+\pi\pi $ \cite{Chao-1} and
$\Upsilon(nS)(n\geq4)\rightarrow \Upsilon(1S)\eta $ \cite{Chao-2},
the transitions $\Upsilon(nS)\rightarrow\eta_b + \gamma$ can occur
via re-scattering sub-processes with the intermediate states being
$B^{(*)}\bar{B}^{(*)}$ where another $B^{(*)}$ meson is exchanged at
the t-channel. The corresponding diagrams are depicted in
Fig. \ref{fig:fsi}. Since the other diagrams can be obtained by a
charge conjugation transformation of $B^{(*)}\rightleftarrows \bar{B}^{ (*)}$ the contribution of each diagram in
Figs. \ref{fig:fsi} (a-f) should be multiplied by a factor 2.

Now let us calculate the amplitudes of the re-scattering processes
which occur at the hadron level. Following Refs.
\cite{Chao-1,Chao-2} we will not account for the contribution from
the dispersive parts of the diagrams where $B^{(*)}$ from
$\Upsilon(ns)$ are off-shell, but only concern the absorptive parts
where $B^{(*)}$ are real particles on their mass-shells. The
off-shell effect of the meson exchanged at t-channel is compensated
by a monepole form factor which is also reflects the inner
structures of the mesons at the effective vertex
\begin{eqnarray}\label{eq:monopole}
\mathcal{F}(m_i,q^2)=\frac{(\Lambda+m_i)^2-m_i^2}{(\Lambda+m_i)^2-q^2},
\end{eqnarray}
where $q$ and $m_i$ are the momentum and mass of the exchanged
meson respectively. And the cutoff is set as $\Lambda=600$ GeV
\cite{Chao-1,Chao-2}.

The absorptive part of the amplitude is read as
\begin{eqnarray}\label{eq:amplitude}
Abs_i=&&\frac{|\mathbf{p}_1|}{32\pi^2m_{\Upsilon(nS)}}\int d \Omega
\mathcal{A}_i[\Upsilon(nS)\rightarrow B^{(*)}\bar{B}^{(*)}]
\nonumber\\&&\times \mathcal{C}_i[B^{(*)}\bar{B}^{(*)}\rightarrow
\eta_b+\gamma]\times\mathcal{F}(m_i,q^2),
\end{eqnarray}
with $i=a,b,c,d,e,f$. Here, $d\Omega$ and $\mathbf{p}_1$ are the
solid angle and linear momentum of the on-shell $B^{(*)}$ in the
rest frame of $\Upsilon(nS)$, respectively.

The effective couplings for $\Upsilon BB$, $\Upsilon\! B^*\!B$ and
$\Upsilon B^*B^*$ we adopt in this work are directly borrowed from
Refs. \cite{Chao-1,Chao-2}, and some discussions will be made in the
last section.
\begin{subequations} \label{effective-Lagrangians1}
\begin{eqnarray}
\mathcal{L}_{\Upsilon BB}&=& g_{\Upsilon
BB}\Upsilon_\mu(\partial^\mu
B{B}^{\dagger}-B\partial^\mu {B}^{\dagger}),\label{L-YBB}\\
\mathcal{L}_{\Upsilon B^*B}&=& \!\frac{g_{\Upsilon\!
B^*\!B}}{m_{\Upsilon}}\varepsilon^{\mu\nu\alpha\beta}\partial_\mu
\!\Upsilon_\nu
\!\nonumber\\
&& \times(B^*_\alpha\overleftrightarrow{\partial}_\beta
{B}^{\dagger}\!\! - \!\!
B\overleftrightarrow{\partial}_\beta{B}^{*\dagger}_\alpha\!),\label{L-YB*B}\\
\mathcal{L}_{\Upsilon B^*B^*}&=& g_{\Upsilon B^* B^*} (
-\Upsilon^\mu
B^{*\nu}\overleftrightarrow{\partial}_\mu {B}_\nu^{*\dagger} \nonumber\\
&&+ \Upsilon^\mu B^{*\nu}\partial_\nu{B}^{*\dagger}_{\mu} -
\Upsilon_\mu\partial_\nu B^{*\mu}
{B}^{*\nu\dagger}),\label{L-YB*B*}
\end{eqnarray}
\end{subequations}
where
$\overleftrightarrow{\partial}=\overrightarrow{\partial}-\overleftarrow{\partial}$
and the coupling constants were fixed as \cite{Chao-1,Chao-2}
\begin{eqnarray}
g_{\Upsilon BB}&&=2.5\nonumber\\g_{\Upsilon B^*B}&&=1.4\pm
0.3\nonumber\\g_{\Upsilon B^*B^*}&&=2.5\pm0.4.\,\label{HQS:g-YBB}
\end{eqnarray}
Following the strategy of Ref.\cite{Chao-2}, we obtain
\begin{subequations} \label{effective-Lagrangians2}
\begin{eqnarray}
\mathcal{L}_{\gamma BB}&=& g_{\gamma BB}A_\mu(\partial^\mu
B{B}^{\dagger}-B\partial^\mu {B}^{\dagger}),\label{L-YBB}\\
\mathcal{L}_{\gamma B^*B}&=& \!\frac{g_{\gamma\!
B^*\!B}}{m_{B^*}}\varepsilon^{\mu\nu\alpha\beta}\partial_\mu
\!A_\nu
\!\nonumber\\
&& \times(B^*_\alpha\overleftrightarrow{\partial}_\beta
{B}^{\dagger}\!\! - \!\!
B\overleftrightarrow{\partial}_\beta{B}^{*\dagger}_\alpha\!),\label{L-YB*B}\\
\mathcal{L}_{\gamma B^*B^*}&=& g_{\gamma B^* B^*} ( -A^\mu
B^{*\nu}\overleftrightarrow{\partial}_\mu {B}_\nu^{*\dagger} \nonumber\\
&&+ A^\mu B^{*\nu}\partial_\nu{B}^{*\dagger}_{\mu} -
A_\mu\partial_\nu B^{*\mu} {B}^{*\nu\dagger}),\label{L-YB*B*}\\
\mathcal{L}_{B^*B\eta_b}&=& ig_{B^*B\eta_b}B^*_{\mu}\partial^\mu\eta_b{B}^{\dagger},\label{L-B*Beta}\\
\mathcal{L}_{B^*B^*\eta_b}&=&
i\frac{g_{B^*B^*\eta_b}}{m_{B^*}}\varepsilon^{\mu\nu\alpha\beta}\partial_{\mu}B^*_{\nu}{B^*}^{\dagger}_{\alpha}
\partial_\beta\eta_b.\label{L-B*B*eta}
\end{eqnarray}
\end{subequations}
With the heavy quark spin symmetry\cite{Isgur:1989vq}, we have
\begin{eqnarray}
g_{\gamma BB}&=&g_{\gamma B^*B}=g_{\gamma B^*B^*},\nonumber\\
g_{\eta_b B^*B}&=&g_{\eta_b B^*B^*}=g_{\Upsilon(1S)
BB}.\,\label{HQS:g-etaBB}
\end{eqnarray}

In terms of the theoretically evaluated value of
$\Gamma(B^{*+}\rightarrow B^+ \gamma)=0.40\pm0.03$
keV \cite{Choi:2007se} one can fix $g_{\gamma B^*B}\approx 3.5$. The
coupling $g_{\eta_b B^*B}$ should be at the order of $O(1)$, but may
vary within a reasonable range. If we choose $g_{\eta_b
B^*B}=g_{\eta_b B^*B^*}\sim 1$ the contributions of  the diagrams
(a), (c), (d), (f) in Fig. \ref{fig:fsi} to
$\mathcal{B}(\Upsilon(5S)\rightarrow\gamma \eta_b)$ are
approximately a few of $10^{-7}$ GeV  and that  of the diagrams (b),
(e) are slightly smaller than the direct transition. If the
interference between the contributions of the direct decay and that
through the re-scattering is constructive, the total width would be
about a few times larger, but if it is destructive, the width would
be very suppressed. However, the coupling $g_{\Upsilon(1S)BB}$ may
be as large as 15 as was estimated in Refs. \cite{Chao-1,Chao-2}. If
we use this value for the coupling constants $g_{\eta_b B^*B}$,
$g_{\eta_b B^*B^*}$ the contributions of  the diagrams in
Fig. \ref{fig:fsi} would enhance the total width by more than two
orders. In principle, the diagram (a) in Fig. \ref{fig:fsi} can
contribute to $\Upsilon(4S)\rightarrow \eta_b+\gamma$, but the mass
of $\Upsilon(4S)$ is just above the threshold of $B^+B^-$, so that
the phase space would greatly suppress the contribution of the
diagram.


\section{Conclusion}

By studying the radiative decay of $\Upsilon(nS)\to
\eta_b+\gamma$, we can learn much about the hadronic structure of
$\eta_b$. Efforts have been made to explore the spin singlet
$\eta_b$, in the Data-book of 2008, $\eta_b$ was still omitted
from the  summary table \cite{PDG08}. In fact, determination of
the mass of $\eta_b$ is made via the radiative decays of
$\Upsilon(nS)\to \eta_b+\gamma$ \cite{:2008vj}, and the recent
data show that $m_{\eta_b}=9388.9^{+3.1}_{-2.3}(stat)\pm
2.7(syst)$ by the $\Upsilon(3S)$ data and
$m_{\eta_b}=9394.2^{+4.8}_{-4.9}(stat)\pm 2.0(syst)$ by the
$\Upsilon(2S)$ data \cite{:2009pz}. Recently, Penin
\cite{Penin:2009wf} reviewed the progress for determining the mass
of $\eta_b$ and indicated that the accurate theoretical prediction
of $m_{\eta_b}$ would be a great challenge. Indeed, determining
the wavefunction of $\eta_b$ would be even more challenging. In
this work, we are not going to obtain the wavefunction or even the
mass of $\eta_b$ based on the fundamental theories, such the
non-perturbative QCD, but using the radiative transition to
testify the phenomenologically determined wavefunction as long as
the mass is well measured. We carefully study the transition rates
of the radiative decays which would help experimentalists to
extract information about $m_{\eta_b}$. The transition rate of
$\Upsilon(1S)\to\eta_b+\gamma$ is very sensitive to the mass
splitting $\Delta M=m_{\Upsilon(1S)}-m_{\eta_b}$ (see the text
above), thus an accurate measurement of the radiative decay may be
more useful to learn the spin dependence of the bottominia.

Recently our experimental colleagues have made some progress. CLEO
and BaBar collaborator measure the
$\Upsilon(3S)\rightarrow\eta_b+\gamma$ and the mass $m_{\eta_b}$
which offer an opportunity for us to more deeply study $\eta_b$.

Following Ref. \cite{Choi:2007se} we systematically study
$\Upsilon(nS)\rightarrow\eta_b+\gamma$ in the LFQM. We take an
alternative way which is different from that adopted in
Ref. \cite{Choi:2007se} to fix the parameters $i.e.$, namely we use the
data to fit $\beta_{\Upsilon(nS)}$ and $\beta_{\eta_b}$ in the
wavefunctions. Then using these parameters we estimate
$\Upsilon(1S)\rightarrow\eta_b+\gamma$,
$\Upsilon(2S)\rightarrow\eta_b+\gamma$,
$\Upsilon(4S)\rightarrow\eta_b+\gamma$ and
$\Upsilon(5S)\rightarrow\eta_b+\gamma$. Our result indicates
$\Upsilon(2S)\rightarrow\eta_b+\gamma$ and
$\Upsilon(5S)\rightarrow\eta_b+\gamma$ are more sensitive to the
parameters than $\Upsilon(1S)\rightarrow\eta_b+\gamma$ and
$\Upsilon(4S)\rightarrow\eta_b+\gamma$.

Since the value of $\Upsilon(1S)\rightarrow\eta_b+\gamma$ is
sensitive to $\Delta M$, we show the dependance of
$B(\Upsilon(1S)\rightarrow\eta_b+\gamma)$ on $\Delta M$. We hope that
our experimental colleagues will conduct accurate measurements in
the near future to determine the precise value of $m_{\eta_b}$.

The anomalous largeness of the branching ratio of
$\Upsilon(5S)\rightarrow\Upsilon(1S,2S)+\pi\pi$ motivates
a hot surf of theoretical studies \cite{Simonov}. It was suggested
that the re-scattering effects may explain the unusual large
branching ratio. This mechanism should be tested somewhere else. In
Ref. \cite{Chao-2} the authors evaluated the effect induced by the
mechanism for
$\Upsilon(5S)\rightarrow\Upsilon(1S,2S)+\eta$ and found
that the corresponding branching ratio is also greatly enhanced
compared to the transition among lower resonances. We suggest to
further test the mechanism at the radiative decays where the
effective electromagnetic vertex is relatively simple. Our result
which is obtained in terms of the LFQM, indicates the branching
ratio of $\Upsilon(5S)\rightarrow\eta_b+\gamma$ is not anomalous
compared to $\Upsilon(mS)\rightarrow\eta_b+\gamma\, (m=1,2,3,4)$  as
long as the re-scattering is not taken into account. However, there
could be a two-order enhancement in magnitude for
$\Upsilon(5S)\rightarrow\eta_b+\gamma$ which is induced by the
re-scattering effects. Thus measurement of
$\Upsilon(5S)\rightarrow\eta_b+\gamma$ would be an ideal probe for
the re-scattering mechanism which successfully explains the data of
$\Upsilon(5S)\rightarrow\Upsilon(1S,2S)+\pi\pi$. This is
one of the tasks of the LHCb which will be operating very soon.

\section*{Acknowledgments}
This project is supported by the National Natural Science Foundation of
China (NSFC) under Contracts Nos. 10705001 and 10775073; the Foundation for
the Author of National Excellent Doctoral Dissertation of P.R. China
(FANEDD) under Contracts No. 200924; the Doctoral Program Foundation of Institutions of
Higher Education of P.R. China under Grant No. 20090211120029; the
Special Grant for the Ph.D. program of Ministry of Eduction of P.R.
China; the Program for New Century Excellent Talents in University (NCET) by Ministry of Education of P.R. China; the Special Grant for New Faculty from
Tianjin University.

\section*{Appendix}

The incoming (outgoing) meson in Fig. \ref{fig:LFQM} has the
momentum ${P}^({'}^)={p_1}^({'}^)+p_2$ where ${p_1}^({'}^)$ and
$p_2$ are the momenta of the off-shell quark and antiquark and
\begin{eqnarray}\label{20}
p^+_1&=&x_1 P^+,\qquad p^+_2 = x_2 P^+,
\nonumber\\
{ p}_{1\perp}&=& x_1{ P}_\perp + { k}_\perp,\qquad { p}_{2\perp}= x_2{
P}_\perp - { k}_\perp,
\nonumber\\
p'^+_1&=&x_1 P^+,\qquad p'^+_2 = x_2 P^+,
\nonumber\\
{ p'}_{1\perp}&=& x_1{ P'}_\perp + { k'}_\perp, \qquad{ p'}_{2\perp}=
x_2{ P'}_\perp - { k'}_\perp\nonumber
\end{eqnarray}
with $x_1+x_2=1$, where $x_i$ and $k_\perp(k'_\perp)$ are internal
variables. $M_0$ and $\tilde {M_0}$ are defined
\begin{eqnarray}\label{app2}
&&M_0^2=\frac{k^2_\perp+m^2_1}{x_1}+\frac{k^2_\perp+m^2_2}{x_2},\nonumber\\&&
\tilde {M_0}=\sqrt{M_0^2-(m_1-m_2)^2}.\nonumber
 \end{eqnarray}

The radial wavefunctions $\phi$ related to $\Upsilon(nS)$ are defined
\begin{eqnarray*}\label{app4}
\phi(1S)&=&4\Big(\frac{\pi}{\beta^2}\Big)^{3/4}\sqrt{\frac{\partial
k_z}{\partial x}}{\exp}\Big(-\frac{k^2_z+k^2_\perp}{2\beta^2}\Big),\nonumber\\
 \phi(2S)&=&4\Big(\frac{\pi}{\beta^2}\Big)^{3/4}\sqrt{\frac{\partial
k_z}{\partial x}}{\exp}\Big(-\frac{k^2_z+k^2_\perp}{2\beta^2}\Big)\nonumber\\&&\times\frac{1}{\sqrt{6}}
\Big(-3+2\frac{k^2_z+k^2_\perp}{\beta^2}\Big),
\nonumber\\
 \phi(3S)&=&4\Big(\frac{\pi}{\beta^2}\Big)^{3/4}\sqrt{\frac{\partial
k_z}{\partial x}}{\exp}\Big(-\frac{k^2_z+k^2_\perp}{2\beta^2}\Big)\nonumber\\&&\times\frac{1}{{2\sqrt{30}}}
\Big(-15-20\frac{k^2_z+k^2_\perp}{\beta^2}+4\frac{(k^2_z+k^2_\perp)^2}{\beta^4}\Big),
\nonumber\\
 \phi(4S)&=&4\Big(\frac{\pi}{\beta^2}\Big)^{3/4}\sqrt{\frac{\partial
k_z}{\partial x}}{\rm
exp}\Big(-\frac{k^2_z+k^2_\perp}{2\beta^2}\Big)\frac{1}{{12\sqrt{35}}}
\nonumber\\&&\times
\Big(-105+210\frac{k^2_z+k^2_\perp}{\beta^2}-84\frac{(k^2_z+k^2_\perp)^2}{\beta^4}\nonumber\\&&+8\frac{(k^2_z+k^2_\perp)^3}{\beta^6}\Big),
\nonumber\\
 \phi(5S)&=&4\Big(\frac{\pi}{\beta^2}\Big)^{3/4}\sqrt{\frac{\partial
k_z}{\partial x}}{\rm
exp}\Big(-\frac{k^2_z+k^2_\perp}{2\beta^2}\Big)\frac{1}{{72\sqrt{70}}}
\nonumber\\&&\times
\Big(945-2520\frac{k^2_z+k^2_\perp}{\beta^2}+1512\frac{(k^2_z+k^2_\perp)^2}{\beta^4}
\nonumber\\&&-288\frac{(k^2_z+k^2_\perp)^3}{\beta^6}+
16\frac{(k^2_z+k^2_\perp)^4}{\beta^8}\Big).
 \end{eqnarray*}
with
\begin{eqnarray}
k_z&=&\frac{x_2M_0}{2}-\frac{m_2^2+k^2_\perp}{2x_2M_0},\\
\frac{\partial k_z}{\partial
x}&=&\frac{M_0}{4x_1x_2}\Big[1-\Big(\frac{m_1^2-m_2^2}{M_0^2}\Big)^2\Big].
\end{eqnarray}
More information can be found in Ref. \cite{Cheng:2003sm}.

\vfil

\end{document}